\documentclass[preprint,preprintnumbers,amsmath,amssymb]{revtex4}
  \makeatletter

\usepackage{graphicx}
\usepackage{dcolumn}
\usepackage{bm}
\newcommand{\ds}{\displaystyle}
\newcommand{\fr}{\frac}
\newcommand{\ra}{\rightarrow}

\newcommand{\G}{\Gamma}

\newcommand{\e}{\epsilon}

\newcommand{\po}{\psi_o}
\newcommand{\poz}{\psi_o(z)}

\newcommand{\pf}{\psi^f}
\newcommand{\pfz}{\psi^f(z)}

\newcommand{\zNz}{Z_N(z)}

\newcommand{\zn}{Z_n}

\newcommand{\zNn}{Z_{N-n}}

\newcommand{\zftz}{Z_t^f(z)}

\newcommand{\re}{\rho_e}
\newcommand{\rlz}{\rho_l(z)}
\newcommand{\rl}{\rho_l}
\newcommand{\q}{\quad}
\newcommand{\qq}{\qquad}

\begin{document}


\title{THEORETICAL STUDY OF COMB--POLYMERS ADSORPTION ON SOLID SURFACES}
\author{Anna Sartori}
\email{sartori@biochem.mpg.de} \altaffiliation[\\{\bf Address for
correspondence}:] { Max Planck Institute for Biochemistry, Dept.
of Molecular Structural Biology, Am Klopferspitz 18a, 82152
Martinsried (Munich), Germany. {\bf Fax n.}: +49 (89) 85782641.}
\author{Albert Johner}
\author{Jean--Louis Viovy}
\author{Jean--Fran\c{c}ois Joanny}
\date{\today}

\vspace{-1cm}

\maketitle

\section*{ABSTRACT}

We propose a theoretical investigation of the physical adsorption
of neutral comb--polymers with an adsorbing skeleton and
non--adsorbing side--chains on a flat surface. Such polymers are
particularly interesting as "dynamic coating" matrices for
bio--separations, especially for DNA sequencing, capillary
electrophoresis and lab--on--chips. Separation performances are
increased by coating the inner surface of the capillaries with
neutral polymers. This method allows to screen the surface
charges, thus to prevent electro--osmosis flow and adhesion of
charged macromolecules ({\it e.g.} proteins) on the capillary
walls. We identify three adsorption regimes: a "mushroom" regime,
in which the coating is formed by strongly adsorbed skeleton loops
and the side--chains anchored on the skeleton are in a swollen
state, a "brush" regime, characterized by a uniform multi--chains
coating with an extended layer of non--adsorbing side--chains and
a non--adsorbed regime. By using a combination of mean field and
scaling approaches, we explicitly derive asymptotic forms for the
monomer concentration profiles, for the adsorption free energy and
for the thickness of the adsorbed layer as a function of the
skeleton and side--chains sizes and of the adsorption parameters.
Moreover, we obtain the scaling laws for the transitions between
the different regimes. These predictions can be checked by
performing experiments aimed at investigating polymer adsorption,
such as Neutron or X-ray Reflectometry, Ellipsometry, Quartz
Microbalance, or Surface Force Apparatus. \\


\section{Introduction}\label{primo}

Polymer adsorption on surfaces is of paramount importance
for numerous applications. In material sciences, it is used to
control surface properties such as wetting, hardness, or
resistance to aggressive environments. It can also have
detrimental effects in fouling, alteration of the aspects of
materials, and generally speaking unwanted changes in surface
properties.

Polymer adsorption also gains more and more attention in the field
of biology, in which it plays an essential role in
bio--compatibility, cell adhesion, containers and instruments
contamination, and in bio--analytical methods. Many proteins, in
particular, have a strong amphiphilic character, and tend to
adsorb easily to surfaces bearing charges or hydrophobic domains.
Strong efforts have been continuously made in the last 20 years,
to develop surface treatments able to prevent unwanted adsorption
of biomolecules in a water environment. The proposed solutions
often amount to treat the surface with specially chosen proteins,
oligomers or polymers. In most cases, a very hydrophilic polymer
is required. Numerous polymers have been proposed in this context,
including polysaccharides, Polyvinyl--alcohol, and the very
popular Polyethyleneoxide.

Depending on the application, two different
approaches can be envisaged: either polymer grafting onto the
surface by one or several covalent bonds (see {\it e.g.}
\cite{Hjerten} etc...), or spontaneous adsorption. The
latter solution is in general easier to implement, and the
regeneration of a fouled or damaged surface is easier. However,
the adsorbed polymer layer is in general more fragile than a
covalently bonded surface, and the adsorption approach puts
constraints on the chemical nature and architecture of the
polymers, that can be difficult to fulfill. In particular, surface
coatings in the field of biology must in most cases be hydrophilic.
Since very hydrophilic polymers most often do not adsorb
spontaneously onto surfaces in a water environment (their free
energy in the solvated state is very low), some "tricks" must be
developed.

One efficient way to favor adsorption of a hydrophilic layer is to
use block copolymers, with one (or several) hydrophilic block, and
one or several blocks playing the role of an anchor. In
particular, diblock and triblock copolymers or oligomers, such as
alkyl--Polyethyleneoxide diblocks, or
Polyethyleneoxide--Polypropyleneoxide--Polyethyleneoxide
triblocks, were used with success for several applications. They
are far to be a universal solution to unwanted adsorption and
wall--interactions of biomolecules. In particular, they seem
rather unsuccessful in the field of capillary electrophoretic
separations.

This bio--analytical method has been widely popularized by the
large genome projects (the human genome project, and most of the
current genome projects rely mainly on capillary array
electrophoresis for DNA sequencing \cite{1}) and its importance is
bound to increase further with the development of massive
"post--genome" screening and "lab--on--chips" methods for research
and diagnosis \cite{5,14}. In capillary electrophoresis, analytes
are separated by electrophoretic migration under a high voltage
(typically 200 to 300 V/cm) in a thin capillary (typically 50
$\mu$m ID). Interactions of the analytes with the walls is of
paramount importance, because it can lead to considerable peak
trailing and because of electro--osmosis, a motion of the fluid
induced by the action of the electric field on the excess of
mobile free charges in the vicinity of a charge surface (see {\it
e.g.} \cite{JLV}).

Consider an infinitely long pipe filled with an electrolyte, with
a non--vanishing Zeta potential (the most common situation when a
solid is in contact with an electrolyte). The surface has a net
charge and a Debye layer of counterions forms in the fluid in the
vicinity of this surface in order to minimize the electrostatic
energy.  In buffers typically used in electrophoresis, the Debye
layer has a thickness of a few nanometers. Choosing for
definiteness a negatively charged surface, such as {\it e.g.}
glass in the presence of water at pH 7, the Debye layer is
positively charged. Applying an electric field along the pipe, the
portion of fluid contained in the Debye layer is dragged towards
the cathode. Solving the Stokes equation with the boundary
condition of zero velocity at the surface leads to a "quasi--plug"
flow profile, with shear localized within the Debye layer and a
uniform velocity in the remainder of the pipe. If the electrical
charge on the surface is non--uniform, due for example to a local
adsorption of biopolymers, the velocity at the wall, which is
imposed by the local zeta potential, is also non--uniform. In such
case, the flow is no more a plug flow, and there are hydrodynamic
recirculations detrimental to the resolution \cite{Ajdari}.
Electro--osmosis generally has very dramatic consequences on the
performance of practical electrophoretic separations, and it must
be thoroughly controlled. Polymers at the interface can play a
double role in circumventing electro--osmosis: by preventing
unwanted adsorption of analytes or impurities contained in the
electrophoretic buffer, and by decoupling the motion at the wall
from the motion of the bulk fluid. For this, the polymer layer
must be thicker than the Debye length. Different strategies, using
either covalently bonded polymers (see {\it e.g.} \cite{16}) or
reversibly adsorbed polymers (so called "dynamic coating", see
{\it e.g.} \cite{17,18}), have been developed. The use of
relatively short copolymers \cite{37,38} has been rather
deceptive, probably because they lead to rather thin layers, and
because the adsorption free energy of an individual chain is too
low to resist the rather aggressive conditions encountered during
electrophoresis in strong fields (in particular high shear at the
wall). These oligomers need to be present in the solution at a
high concentration, in order to yield sufficient dynamic coating.
Presently, the most efficient applications of dynamic coating have
involved Poly--Dimethyl Acrylamide (PDMA) \cite{12} or copolymers
of this polymer with other acrylic monomers \cite{24}. This
polymer seems to present an interesting affinity to silica walls,
thanks to the presence of hydrogen bonding, while remaining
soluble enough in water to behave as a sieving matrix. It was
demonstrated \cite{22}, however, that its reduced hydrophilicity
as compared with, {\it e.g.} Acrylamide \cite{6}, results in
poorer separation performance, probably due to increased
interactions with the analytes.

Recently, we proposed \cite{26} a new family of block copolymers
comprising a very hydrophilic Poly--Acrylamide (PA) skeleton and
PDMA side--chains, as dynamic coating sieving matrices for DNA
electrophoresis. These matrices provide an electro--osmosis
control comparable to that of pure PDMA, while allowing for better
sieving. A large range of different microstructures can be
conceived and constructed, varying the length and chemical nature
of the grafts and of the skeleton, and the number of grafts per
chain. The aim of the present article is to investigate
theoretically the adsorption mechanisms and the structure of
adsorbed layers of polymers with this type of microstructure, in
order to better understand the properties, and to provide a
rational basis for further experimental investigations and
applications. The adsorption of homopolymers, of di- or triblock
copolymers, and random copolymers of adsorbing and non--adsorbing
polymers, has been investigated theoretically
\cite{JoaJoh1,SeBoJohJoa,SeJoaJoh,MaJoa1,JohJoa,DeGads1,DeGads2,MaJoa,MaJoaLei}.
To our knowledge, however, the case of a comb--polymer with
different adsorption properties on the skeleton and on the grafts
has never been considered. We address this problem here,
generalizing on previous work on random and triblock copolymers.

The architecture of the paper is as follows: In Section
\ref{secondo} we investigate the adsorption of comb--polymers with
adsorbing backbone and non-adsorbing side--chains. In Section
\ref{secondo1} we present a mean field model for the analysis of
the adsorption of combs in the limit of small side--chains, {\it
i.e.} in the mushroom regime ($N_t \ll N_B$ , where $N_t$ and
$N_B$ are the number of monomers of a non--adsorbing side--chain
and of the corresponding adsorbing backbone chain section,
respectively).

In Section \ref{secondo2}, we use a scaling approach to describe
the adsorption of comb--polymers both in the mushroom and in the
brush regime ({\it i.e.} in the limit of large side--chains) and
the cross--over between these two regimes.

\newpage

\section{Adsorption of Comb--Polymers with Adsorbing Backbone
and non--Adsorbing Side--Chains: Mean Field Theory
}\label{secondo}

In this section we give a mean field description of the adsorption
of comb--polymers  with adsorbing backbone and non--adsorbing
side--chains onto a flat solid substrate. The combs have an
adsorbing backbone B, made of p blocks of $N_B$ monomers each with
gyration radius $R_{B} \sim a\,N_B^{1/2}$, on which are grafted
$p$ non--adsorbing T side--chains, of $N_t$ monomers each and of
gyration radius $R_t\sim a\,N_t^{1/2}$, where $a$ is the monomer
size. We study the adsorption behavior of combs, which is
monitored by the mass ratio between the adsorbed backbone B and
the non--adsorbing side--chains T.

\begin{figure}[ht]
\vspace{.2in} \centerline {
\includegraphics[width=0.4\textwidth]{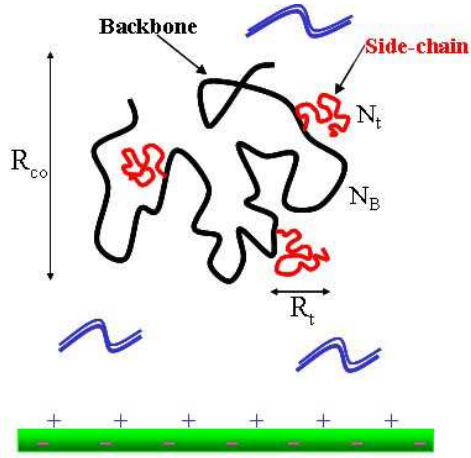}
} \caption{\label{fig:1} {\small Comb--copolymer in solution, made
of an adsorbing backbone made of $p$ blocks of $N_B$ monomers each
and with $p$ non--adsorbing side--chains of $N_t$ monomers each.
The total number of monomers is $N=p\,(N_B+N_t)$ (with $p =3$),
and the gyration radius $R_{\rm co}\sim a \,N^{1/2}$, $a$ being
the monomer size.}}
\end{figure}

\begin{figure}[ht]
\vspace{.2in}
\centerline {
\includegraphics[width=0.4\textwidth]{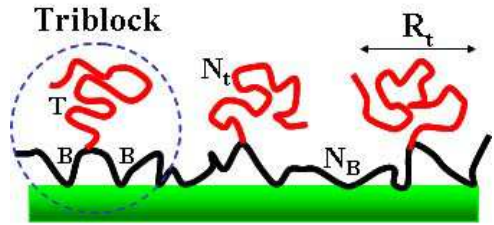}
} \caption{\label{fig:2} {\small Mushroom configuration for a
comb--polymer, made out of $p$ ($p=3$) triblock sub--units. Each
triblock is made of $2$ backbone blocks of $N_B/2$ monomers and of
one anchored side--chain of $N_t$ monomers and gyration radius
$R_t$ (indicated in the Figure with the letters $B$ and $T$,
respectively).}}
\end{figure}

We treat at the mean field level the case of small side--chains
and low grafting density. This leads to a ``mushroom''
configuration for the adsorbed comb--polymers, having the backbone
adsorbed on the surface and the side--chains dangling from the
backbone, with weak steric interaction between them (see
Fig.\ref{fig:1}). Our analysis is based on the idea of describing
the combs as linear chains having 'triblock copolymers' as
renormalized monomers (see Fig.\ref{fig:2}). By exploiting the
known solution for linear chains adsorbing from a dilute solution
on a flat surface, one can directly derive the comb--polymers
adsorption profile, by integrating out the triblock copolymers
degrees of freedom.

\subsection{Mean Field Theory of Triblock Copolymer Adsorption}
\label{secondo1}

Consider a dilute solution of triblock copolymers (p=1), with an
adsorbing backbone made of two large adsorbing blocks of $N_B/2$
monomers each, and one non--adsorbing small side--chain, of $N_t$
monomers, with $N_B \gg N_t$. The total number of monomers is
 $N=N_B+N_t \sim N_B$. We solve the problem in the
Ground State Dominance Approximation, applicable to the case where
the polymers have large enough molecular weights and there is only
one bound state of energy $E_o<0$. A detailed description of the
triblock copolymer adsorption behavior and configurations can be
obtained by deriving the polymer partition function $Z_N(z)$, in
the case where one backbone end--point is fixed at position $z$
above the adsorbing surface ($z=0$) and the other end--point is
free. The partition function $Z_n(z)$ is a solution of the
so--called Edwards equation \cite{SeJoaJoh,JoaJoh1}:
\begin{equation}\label{Edw1}
\left\{
\begin{array}{lll}
-\ds{\fr{\partial Z_n(z)}{\partial n}}&=&-\ds{\fr{\partial^2
Z_n(z)}{\partial z^2}}\, +\,U(z)\,\, Z_n(z),  \\ \\
\left.-\ds{\fr{\partial \ln Z_n(z)}{\partial
z}}\right|_{z=0}&=&-\ds{\fr{1}{b}},
\end{array}
\right.
\end{equation}
where the unit length has been defined as $a/\sqrt{6}$ and $a$ is
the monomer size. The boundary condition imposed at the surface is
a good approximation of the surface potential effect, if we
neglect the details of the concentration profile close to the
surface. The extrapolation length $b$ gives a measure of the
strength of the adsorption. We build here the partition function
of the adsorbed combs starting from the known solution
$Z^o_{N}(z)$ for the adsorption of a linear chain of $N$ monomers,
which is also a solution of Eq.(\ref{Edw1}). Far from the surface
the chain is free and can assume all configurations in space,
$Z^o_{N}(z\gg R_B)=1$. We choose the normalization of the
partition function to be one when the chain reduces to one
monomer, $Z^o_0(z)=1$.

The mean field effective potential experienced by the adsorbed
chains can be expressed self--consistently as:
\begin{equation}\label{U}
U(z) \;=\; \Phi(z) \, - \, \Phi_b \;=\; v\,(c(z)-c_b),
\end{equation}
where $\Phi(z)$ is the monomer volume fraction, equal to the bulk
value $\Phi_b$ sufficiently far from the surface, $v$ is the
excluded volume parameter ($v\sim a^3>0$ in a good solvent) and
$c(z)$ is the monomer concentration.

The partition function $Z^o_N$ can be split into two
contributions: the adsorbed states contribution $Z_N^{o^{a}}$,
arising from chains having at least one monomer adsorbed onto the
surface, and a free chain contribution $Z_N^{o^{f}}$, arising from
chains with no adsorbed monomers, so that $Z^o_N = Z_N^{o^{a}} +
Z_N^{o^{f}}$. We can derive an expression for both components as
independent solutions of the Edwards equation, with appropriate
boundary conditions (see Section \ref{secondo11}). Starting from
the knowledge of $Z_N^{o^a}$ and of $Z_N^{o^{f}}$ (which from now
on we denote as $Z_N^{o}$ and $Z_N^f$) for the adsorption of a
linear chain of $N$ monomers, we build the partition function of a
triblock copolymer, by neglecting the effect of free triblock
copolymer chains on the adsorption profile near the surface. The
non--adsorbing side--chains contribute as a small perturbation to
the adsorption profile.

\subsection{Backbone Partition Function}\label{secondo11}

\begin{figure}[ht]
\vspace{.2in}
\centerline {
\includegraphics[width=0.8\textwidth]
{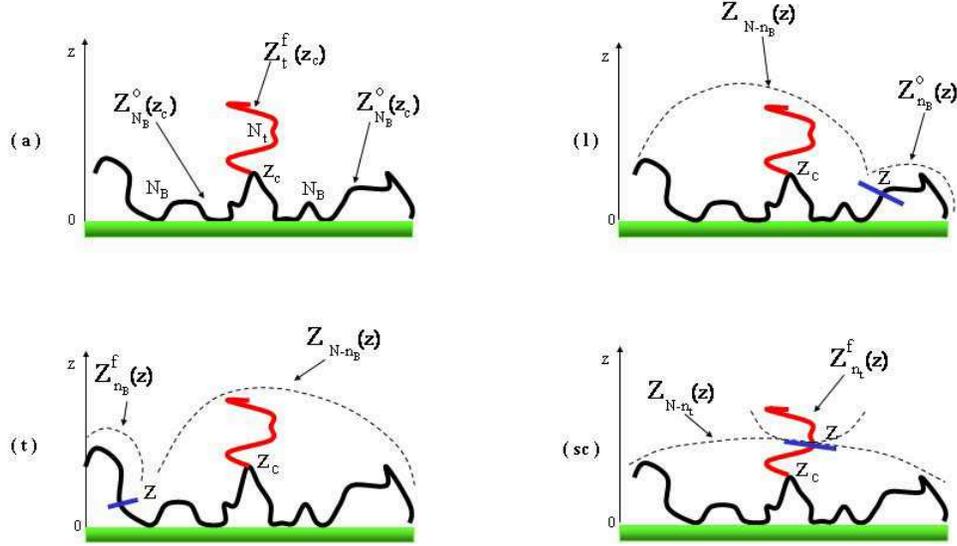}}\caption{\label{fig:3} {\small  Representation of the
construction of the triblock total partition function $Z$ ({\bf
a}) and of the ({\bf l}) backbone's loops ({\bf l}) tails ({\bf
t}) and side--chains ({\bf sc}) contributions to the triblock
copolymer monomer volume fraction $\Phi^{\rm TB}$, starting from
the knowledge of the partition functions for the adsorbed backbone
blocks, $Z^o$ and for the dangling tails, $Z^f_t$ ($z_c$ is the
triblock's core position above the adsorbing surface).}}
\end{figure}

For the adsorption of a linear backbone, made of $N_B$ monomers,
the Ground State Dominance Approximation amounts to considering
the limit of very large molecular weights, $\e_o\,N_B \gg 1$,
where $\e_o$ is the absolute value of the contact free energy per
monomer ($\e_o=|E_0|=-E_0$). This implies that the expansion of
$Z_{N_B}^{o}(z)$ in terms of the normalized eigenvectors
$\psi_i(z)$ ($ \int_0^{\infty} \;{\rm d}z\, \psi_i \, \psi_j \,=\,
\delta_{ij}$) and of the eigenvalues of Eq.(\ref{Edw1}) is
dominated by the first ground state term:
\begin{equation}\label{Zads}
Z_{N_B}^{o}(z)\quad \sim \quad \sum_i \, k_i \; \psi_i(z) \,
e^{\e_i\,N_B}\quad \sim \quad k_o \; \poz\,e^{\e_o\,N_B}\,+
\,...\qquad \e_o\,N_B \gg 1,
\end{equation}
where $\psi_o$ is a solution of:
\begin{equation}\label{psieq}
\begin{array}{lll}
0              &=&  -\ds{\fr{\partial^2 \poz}{\partial z^2}}\,
+\,(U(z)+\e_o)\,\, \poz,
\end{array}
\end{equation}
with boundary condition:
\begin{equation}\label{psieqbc}
\begin{array}{lll}
\ds{\fr{-1}{b}} &=&  \left. \ds{\fr{1}{\po}\, \fr{\partial
\po}{\partial z}}\right|_{z=0} \qquad {\rm and} \qquad \po(z \ra
\infty)\,=\,0.
\end{array}
\end{equation}
The amplitude $k_o$ is fixed by imposing the backbone end--points
conservation relation (see also \cite{Grosberg}):
\begin{equation}\label{ec}
\ds{\fr{\G_o}{N_B}}\q=\q\ds{\int_0^{\infty}}\,\ds{\fr{\rho_{e}(z)}{2}}\;{\rm
d}z,
\end{equation}
where the parameter $\G_o$ represents the surface coverage (number
of monomers per unit surface) and $\rho_e(z)$ is the end--points
monomer density. Close to the surface, the end--points density
$\rho_e(z)$ is proportional to the total partition function of the
linear backbone, $\re(z) \simeq 2\; \Phi_b \; Z^o_{N_B}(z)/N_B$.
The adsorbance $\G_o$ is directly obtained by integrating the
monomer volume fraction $\Phi(z)$, which can be expressed as the
sum of the loops and of the dangling tails contribution,  $
\Phi(z)= \Phi_{\ell}(z) \,+\, \Phi_t(z)$, with:
\begin{equation}
\Phi_{\ell}(z) \,=\,\ds{\fr{\Phi_b}{N_B}} \, \ds{\int}
Z^o_{N_B-n}(z)\,Z^o_{n}(z)\,dn,
\end{equation}
giving $k_o=\int \,\psi_o$, $\Phi_{\ell} \sim \G_o \, \psi_o^2$
and $\G_o = \Phi_B \, k_o^2 \, e^{\e_o\,N_B}$. \\
By taking into account the contributions of the free dangling
backbone ends, one can define the order parameter $\varphi(z)\;
\sim \; \int \; Z^f_{n_B}(z) \; e^{-\e_o\,n_B} \; dn_B,$ which is
solution of the following equation:
\begin{equation}\label{phieq}
\begin{array}{lll}
1      &=& -\ds{\fr{\partial^2 \varphi(z)}{\partial
z^2}}\,+\,(U(z)+\e_o)\,\, \varphi(z),
\end{array}
\end{equation}
derived from Eq.(\ref{Edw1}) for $Z_{n_B}^f(z)$, with boundary
condition $\varphi(0)\,=\,0$. One can solve the
differential equations for $\poz$ and $\varphi(z)$ and express the
effective potential $U$ (or equivalently the monomers volume
fraction $\Phi$ in the case of dilute solutions) as:
\begin{equation}
\Phi \quad = \quad \Phi_{\ell} \,+\, \Phi_t \q=\q \G_o \, \po^2
\,+\, B_o \, (\sqrt{\G_o}\,\po) \, \varphi,
\end{equation}
where:
\begin{equation}\label{po}
\begin{array}{lll}
\sqrt{\G_o}\,\po &=&\left\{
\begin{array}{l}
\sqrt{2}\q (z+b)^{-1} \qquad 0 < z < z^*\\
\sqrt{2} \; (N_B\,b) \q z^{-4} \qquad z^* < z < \lambda_o,
\end{array}
\right. \\ \\
\varphi &=&\left\{
\begin{array}{l}
z^2/3 \;\log(z/z^*) \qquad 0 < z < z^*\\
z^2/18  \qquad z^* < z < \lambda_o,
\end{array}
\right.
\end{array}
\end{equation}
with $z^* \sim (N_B\,b \cdot \ln(N_B/b^2))^{1/3}$, $\G_o\sim2/b$,
$k_o = \sqrt{2/\G_o}\; \log(z^*/b) \sim 1/3 \,\sqrt{b}\;
\log(N_B/b^2)$ and $B_o = 2\,\sqrt{\G_o}/(N_B\,k_o)$. The value
$\e_o$ of the ground state energy can be expressed as:
\begin{equation}
\begin{array}{lll}
\e_o & = & \ds{\fr{1}{N_B}} \; \ln \left[
\ds{\fr{\G_o}{\Phi_B\;k_o^2}} \right]\q \sim \q \ds{\fr{1}{N_B}}
\; \ln \left[ \ds{\fr{2}{\Phi_B\,b^2}} \right],
\end{array}
\end{equation}
The parameter $\lambda_o \sim \epsilon_o^{-1/2}$ is the average
thickness of the adsorbed layer. \\ \\
The non--adsorbing side--chain states are described by the ``free
states'' function $Z^f_{N_t}(z)=\zftz$ (with $Z^f_t(z=0)=0$ and
$Z^f_t(z \rightarrow\infty) =1$) of a linear non--adsorbed chain
of $N_t$ monomers, with the constraint of having one end--point
anchored at the middle of the adsorbing backbone at $z=z_c$, the
other end point being integrated out (see Fig.\ref{fig:3}). The
non--adsorbing side--chains feel a strong entropic repulsion at
the surface within a layer of thickness $R_t\,\sim \, a\,
N_t^{1/2}$, where $R_t$ is their radius of  gyration. Thus, for $0
\,\leq z \leq \,R_t$, the probability of finding ``free''
side--chains is very low, $\zftz \ll 1$, while for $z>R_t$ one has
$\zftz \sim 1$. A detailed calculation of the propagator and of
the partition function of free chains is given in Appendix C
assuming that the side chains only give a small contribution to
the total concentration profile; for simplicity, we only give here
scaling arguments.

Close to the surface, the most important contribution to the
adsorption profile comes from the central monomers of the
side--chains \cite{SeJoaJoh}, meaning that $\zftz \sim \pfz$,
where $\pfz$ describes 'free' (non--adsorbing) states and
satisfies the same equation as $\poz$, but with different boundary
conditions at the surface, $\pf(0) \,=\, 0$. For $0\leq z \leq
R_t$, one can express $\pfz$ as a function of $\poz \sim 1/z$,
$\pfz\,=\,\po\,\ds{\int_{d}^{z}}\,\po^{-2}\;{\rm d}z\,\sim\,z^2$,
so that:
\begin{equation}\label{pftztot}
\pfz \,\sim\, \left\{
\begin{array}{ll}
\ds{\fr{z^2}{R_t^2}} & z \leq R_t,  \\
1         &   R_t<z\le \lambda_o.
\end{array}
\right.
\end{equation}

\subsection{Triblock Copolymer Partition Function}\label{secondo12}

We now derive the triblock copolymer partition function $Z_N(z)
\sim k' \, \poz \, e^{\epsilon\,N_B}$ (where $k'$ is a constant to
be determined and $\e$ is the adsorption energy per monomer of the
triblock, with $\epsilon\,>\,\epsilon_o$) by proceeding
analogously to the case of a linear chain and taking into account
the constraint of having a non--adsorbing side--chain anchored at
the mid--point of the adsorbing backbone. We start by deriving the
total partition function for a triblock copolymer, which is built
from the knowledge of the linear backbone partition function
$Z^o_{N_B/2}$ and the partition function of the anchored
side--chains $Z^f_t$:
\begin{equation}
Z \quad = \quad \int_{0}^{+\infty}  \,  {\rm d}z_c \,
\int_{0}^{+\infty} \, {\rm d}z \;\; Z^o_{N_B/2}(z,z_c) \;
Z^f_t(z_{c})\; Z^o_{N_B/2}(z_c) \; = \; Z^o \;\;
\ds{\fr{e^{(\epsilon-\epsilon_o)\,N_B}}{\Gamma_o\,R_t}},
\end{equation}
where $z_c$ is the vertical coordinate of the core ({\it i.e.} the
side--chain anchoring point) above the surface (see
Fig.\ref{fig:2}), $Z^o_{N_B/2}(z,z_c)=\psi_o(z) \, \psi_o(z_c) \,
e^{\e\,N_B/2}$ is the partition function of an adsorbing backbone
block with one end at $z_c$ and one end at $z$, $Z^f_t(z_{c})$ is
the partition function of the non--adsorbing side--chains with one
end at $z=z_c$ and $Z^o= k_o^2 \, e^{\epsilon_o\,N_B}$. By
combining the classical chemical potential balance for a triblock and a
linear chain ($\Phi_b/N=\Gamma/(N \,Z)$) one obtains:
\begin{equation}
\ln\left(\fr{\Gamma}{\Gamma_o}\right)-N_B\,(\epsilon-\epsilon_o)\,+\,\ln(\Gamma_o\,R_t)
\,=\,0
\end{equation}
Assuming again that the side--chains give only a small
perturbation to the total concentration in the adsorbed layer, one can approximate the
triblock copolymer surface coverage $\Gamma$ by the linear chain surface
coverage, $\Gamma \sim \Gamma_o$, leading to the following
expression for the triblock copolymer ground state energy:
\begin{equation}
\epsilon \quad = \quad
\epsilon_o\,+\,\ds{\fr{1}{N_B}}\ln(\Gamma_o\,R_t).
\end{equation}
The density of junction points of a triblock copolymer at position
$z_c$ above the adsorbing surface can be built using similar
arguments to those used for the total partition function. Starting
from the single chain partition functions $Z^f_t$ and
$Z^o_{N_B/2}$, $\rho^{\rm TB}_{\rm c}(z_c)\, = \, \Phi_b
\,Z^f_t(z_c) \cdot Z^{o^2}_{N_B/2}(z_c)/N$, one gets:
\begin{equation}\label{cores}
\rho^{\rm TB}_{\rm c}(z_c) \q=\q \left\{
\begin{array}{lll}
\ds{\fr{\Gamma_o}{N_B\,R_t}}  & \q &\q z_c<R_t,\\ \\
\ds{\fr{\Gamma_o\,R_t}{N_B}} & \q z_c^{-2}  & \q R_t<z_c<z^* \\
\\
\Gamma_o \, N_B \, R_t& \q z_c^{-8} &\q z^*<z_c<\lambda,
\end{array}
\right.
\end{equation}
with $\int \rho^{\rm TB}_c(z) \, {\rm d}z = \Gamma_o/N_B$. \\
The junctions of the adsorbed triblock copolymers are therefore
confined close to the surface; as they must belong to the loops of
the backbone chains, their density in the region where the
concentration is dominated by monomers belonging to tails is
extremely low. Our analysis holds as long as the gyration radius
of the side--chains $R_t$ (which gives a measure of the thickness
of a depletion layer close to the surface) remains smaller than
$z^*$, {\it i.e.} for:
\begin{equation}\label{mb}
N_t \leq N_B^{2/3}.
\end{equation}
For $N_t \geq N_B^{2/3}$ the surface coverage is dominated by the
side--chains contribution and the side--chains gyration radius
becomes comparable to the thickness of the loops adsorbed layer,
$R_t \sim z^*$. This means that the side--chains contribution to
the monomer volume fraction can no longer  be treated as a
perturbation. We can infer that this condition identifies the
cross--over from ``mushroom'' ($N_t < N_B^{2/3}$) to ``brush''
($N_t
> N_B^{2/3}$) configuration for the adsorbed triblock copolymers.
A more rigorous derivation of this statement is obtained at the
end of Section \ref{secondo14}.

\subsection{Monomer Volume Fraction}\label{secondo13}

\begin{figure}[ht]\hspace{0cm}
\vspace{.2in}
\centerline {
\includegraphics[width=0.5\textwidth]
{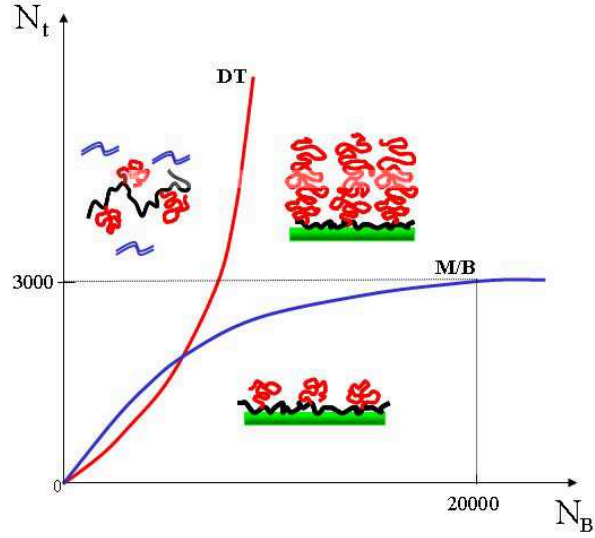}}\caption{\label{fig:5} {\small Phase diagram for a
comb--polymer with an adsorbing backbone made of $p$ blocks of
$N_B$ monomers each and non--adsorbing side--chains of $N_t$
monomers each. The dashed curve represents the mushroom to brush
configuration threshold (M/B) while the solid curve represents the
desorption threshold (DT).}}
\end{figure}

In order to get the monomer concentration profiles, we need to evaluate the
effective potential felt by the adsorbed chains. For sufficiently
diluted solutions, $\Phi_b$ can be neglected and thus $U(z) \sim
\Phi(z)$. The triblock copolymer partition function $\zNz$ can be
expressed by splitting the chain into two parts and by summing over
all possible distinct configurations (see Fig.\ref{fig:3}):
\begin{equation}\label{fitot}
\Phi(z) \; =\; \ds{\fr{\Phi_b}{N_B}} \, \sum\, \int \, dn
 \;\; \zNn \; \zn
\; \sim \; \Phi_{\ell}(z) \,+\, \Phi_t(z)  \,+\, \Phi_{\rm sc}(z),
\end{equation}
where '$\ell$', 't' and 'sc' stand for loops, tails and
side--chains contribution, respectively and:
\begin{equation}\label{fi}
\begin{array}{lll}
\Phi_{\ell}(z) &=& \ds{\fr{\Phi_b}{N_B}}\;\ds{\int_o^{N_B}} dn_B
\ds{\int_o^{+\infty}} {\rm d}z_c\; Z^o(z_c) \; Z_t^f(z_c) \;
Z^o(z,z_c) \; Z^o(z)\;=\;\Gamma_o \;
\psi_o^2(z),\\ \\
\Phi_{t}(z) &=& \ds{\fr{2\,\Phi_b}{N_B}}\;\ds{\int_o^{N_B}} dn_B
\ds{\int_o^{+\infty}} {\rm d}z_c\; Z^o(z_c) \; Z_t^f(z_c) \;
Z^o(z,z_c) \; Z^f_{n_B}(z)\;=\;\ds{\fr{2\,\Gamma_o}{N_B\,k_o}}\;
\psi_o(z)\;\varphi(z),\\ \\
\Phi_{\rm sc}(z) &=& \ds{\fr{\Phi_b}{N_B}}\;\ds{\int_o^{N_t}}
dn_t \;\ds{\int_{0}^{\infty}} {\rm d}z_c\; Z_{N/2}^{o}(z_c)\;Z_{N/2}^{o}(z_c)\;
Z_t^f(z,n_t\vert z_c)  \; Z_t^f(z,N_t-n_t),
\end{array}
\end{equation}
where $Z_t^f(z_\omega,n\vert z_\alpha)$ is the partition function
of an non--adsorbed strand of $n$ monomers starting at $z_\alpha$
and ending at $z_\omega$, $Z^f_t(z_\omega,n)$ is obtained after
integration of $Z_t^f(z_\omega,n\vert z_\alpha)$ over  $z_\alpha$.
These partition functions are discussed in Appendix C. The
expressions (\ref{fi}) for the loops and tails contributions to
the total monomer volume fraction for a triblock copolymer of
$N\sim N_B$ monomers are the same as the one found for the
adsorption of a linear polymer chain of $N_B$ monomers. For the
side--chains monomer volume fraction $\Phi^{\rm TB}_{\rm sc}(z)$,
one finds (see Fig.\ref{fig:5}):
\begin{equation}\label{fisc}
\Phi^{\rm TB}_{\rm sc}(z) \q=\q \left\{
\begin{array}{lll}
\ds{\fr{\Gamma_o}{N_B\,R_t}} & \q z^2   & \q d<z<R_t,\\
\ds{\fr{\Gamma_o\,R_t^3}{N_B}} & \q z^{-2}  & \q R_t<z<z^*,\\
\Gamma_o\,R_t^3\,N_B  & \q z^{-8} & \q z^*<z<\lambda,
\end{array}
\right.
\end{equation}
where $\lambda$ is the average thickness of the triblock copolymer
adsorbed layer. Close to the wall, $d<z<R_t$, where the side
chains are depleted, the side--chain volume fraction is
constructed from chemically close junction points, each of which
contributes $z^2$ monomers (one blob). Further from the wall, the
side--chain density follows the junction point density (with the
normalization factor $N_t$). At short distances, in the adsorbed
loops layer, $\Phi_{\ell}(z)$ dominates over the other
contributions, at large distances, $\Phi(z) \sim \Phi_t(z)$. As
for linear chain adsorption, the characteristic length $z^*$
represents the length--scale inside the adsorbed layer (where the
adsorption energy $\e$ is negligible compared to the effective
potential $U(z)$), at the cross--over from the loops--dominated
layer to the tails dominated layer:
\begin{equation}\label{zstar2}
\begin{array}{lll}
z^* &\sim& \left[ N \, b\; \ln{\left(\ds{\fr{N_B}{b^2}}\right)}
\right]^{1/3} \;\sim\; (N_B\,b)^{1/3},
\end{array}
\end{equation}
which is the same as the one found for the adsorption of a linear
chain of $N_B$ monomers.

The thickness of the adsorbed triblock copolymer layer
$\lambda\sim \epsilon^{-1/2}$ is smaller than the
thickness of a linear polymer adsorbed layer:
\begin{equation}
\label{lambda}
\lambda \q  = \q \lambda_o \left[1\,+\, \lambda_o^2
\,\ds{\fr{\ln{(\Gamma_o\,R_t)}}{N_B}}\right]^{-1/2},
\end{equation}
where
\begin{equation}
\lambda_o \q  = \q \left(\fr{N_B\,a^2}{6}\right)^{1/2} \,
\ln{ \left(\fr{1}{\Phi_b\,b^2}\right)}^{-1/2}.
\end{equation}
If $N_t$ increases, $\lambda$ decreases, as the adsorption is
partially prevented by the presence of the side--chains.\\
From the complete expression for the triblock copolymer volume
fraction $\Phi(z)$ (see Eq.(\ref{fitot}),(\ref{fi}) and
(\ref{fisc})), one can calculate the correction to the surface
coverage (of backbone monomers) due to the side--chains:
\begin{equation}
\Gamma \q= \q \Gamma_o \;\left[1-\fr{2}{\Gamma_o\,\lambda}
\;+\;\fr{N_t}{N_B} \,\left( 2\,-\,\fr{R_t}{z^*} \right)\right].
\end{equation}

\subsubsection{From Triblock Copolymers to Combs}\label{secondo14}

Once the adsorption profiles in triblock copolymer adsorbed layers
are known, it is straightforward to extend the study  to
comb--like architectures. The combs are linear polymers made of
$p$ sub--units ($p \ge 1$), which are triblock copolymers with an
adsorbing backbone of $N_B$ monomers and size $R_B=N^{1/2}a$ and
one non--adsorbing side--chain of $N_t$ monomers, with $N_t \ll
N_B$. The total number of monomers per comb is $N \,=\, p\,(N_B
+N_t) \sim p\,N_B$.\\
Far from the adsorbing surface, {\it i.e.} for $z \gg R_B$, the
comb--polymer can be seen as a chain of blobs, each being a
triblock copolymer. The density of cores (branching points) and
the volume fraction of side chain monomers are therefore
$\rho_c(z) = 2 \, z^{-2}/N_B$ and $\Phi_{\rm sc}(z)=2\,N_t \,
z^{-2}/N_B$. Close to the surface, Eq.(\ref{fisc}) for triblock
copolymer adsorption correctly predicts the structure of the comb
side--chains adsorption profile $\Phi_{\rm
sc}(z)={\fr{\Gamma_o\,R_t^3}{N_B}} z^{-2}$. These two predictions
crossover smoothly. There is therefore an intermediate regime
involving a new length scale. There is actually a strong
constraint that loops smaller than $N_B$ monomers cannot contain
more than one branching point. The profile given by
Eq.(\ref{fisc}) is thus valid only up to a distance $z_1$ where
each loop comprises a number of side--chains of order one.  One
can estimate the fraction $x$ of loops of size $z$ that contain
one side--chain, $x=\rho_c/\rl$, where the loops density $\rl$ is
given by the monomer density divided by the number of monomers per
loop $g(z) \sim z^2$ (Gaussian loops), $\rlz \sim z^{-4}$. This
fraction is smaller than one  if $z<z_1$, where
\begin{equation}
z_1 \q=\q \left( \ds{\fr{N_B}{N_t^{1/2}}}  \right)^{1/2},
\end{equation}At distances $z_1 < z < R_B$, all loops contain a branching point and
$x\sim 1$, {\it i.e.} $\rho_c(z) \sim \rlz$. This description
holds as long as the adsorbed combs are in a mushroom regime, {\it
i.e.} for $R_t<z_1<R_B$, which leads to the same threshold derived
at the end of Section \ref{secondo12} for the crossover from a
mushroom to a  brush configuration for the adsorbed triblock
copolymers:
\begin{equation}
N_t<N_B^{2/3}.
\end{equation}
The structure of the adsorbed layer is mainly determined by the
small loop structure close to the wall which is the same for comb
and triblock copolymers. In particular, the monomer chemical
potential $\epsilon$ is the same in both cases. This gives the
adsorbed layer thickness of a comb--polymers comprising $p$
blocks, $\lambda_p=\sqrt p\lambda$ where $\lambda$ is given by
Eq.(\ref{lambda}). In Table \ref{coo} we present a summary of the
mean field behavior of the comb--polymers branching point density
and monomer volume fraction in the adsorbed layer.
\begin{table}[h]\label{coo}
\begin{center}
\begin{tabular}{|c||c|c|c|c|}
\hline & \hspace{0.1cm}$0 \leq z \leq R_t$ \hspace{0.1cm}&
\hspace{0.1cm} $R_t \leq z \leq z_1$ \hspace{0.1cm}&
\hspace{0.1cm} $z_1 \leq z \leq R_B $ \hspace{0.1cm}&
\hspace{0.1cm} $R_B \leq z \leq \lambda_p $ \hspace{0.1cm}\\
 \hline
$\rho_c(z)$            &  $\fr{\Gamma_o}{N_B\,R_t}$ &
$\fr{\Gamma_o\,R_t}{N_B} \; z^{-2}$ & $\Gamma_o\,z^{-4}$ &
                        $\fr{\Gamma_o}{N_B}\; z^{-2}$ \\
\hline
 $\Phi^{\rm co}_{\rm sc}(z)$  &
$\fr{\Gamma_o}{N_B\,R_t} \;z^2$ &
$\fr{\Gamma_o\,R_t^3}{N_B}\;z^{-2}$ & $\Gamma_o\,N_t\;z^{-4}$ &
$\fr{\Gamma_o\,N_t}{N_B}\; z^{-2}$\\
\hline
\end{tabular}
\caption{\label{coo} Mean field branching point density
$\rho_c(z)$ and monomer volume fraction $\Phi_{\rm sc}(z)$ for an
adsorbed comb--polymer (z is the vertical coordinate above the
adsorbing surface).}
\end{center}
\end{table}

\section{Comb--Copolymer Adsorption: Scaling Approach} \label{secondo2}

\subsection{Comb--Copolymer Mushroom Regime: Scaling}
\label{secondo21}

We now generalize our description of the adsorption of a
comb--polymer in a mushroom configuration ($R_t < z_1$) using a
scaling approach. As in the mean field theory, the side--chains
give only a very small perturbation to the concentration profile
and the total monomer concentration profile decays with the same
power law as that of adsorbed linear polymer chains  $\Phi(z) \sim
z^{1/\nu-d}$, where $\nu\approx 0.59$ is the Flory scaling
exponent and $d=3$ the space dimension \cite{DeGads2}. For each of
the regimes found in the mean field theory, we now derive the
corresponding scaling laws.

At distances from the adsorbing surface larger than the Flory
radius of the side chains $R_t\sim a\, N_t^{\nu}$, the
side--chains behave essentially as free chains. As in the mean
field theory, if $R_t<z<z_1$, the density of side--chains is
proportional to the total monomer density  and $\Phi_{\rm sc}(z) =
c_1 \cdot z^{1/\nu-d}$. The constant $c_1$ is determined by
imposing the conservation of the side--chain monomers, $\int
\Phi_{\rm sc}(z) \, {\rm d}z = \Gamma\,N_t/N_B$, leading to:
\begin{eqnarray}
\Phi_{\rm sc}(z) \q &=& \q \ds{\fr{\Gamma\,N_t}{N_B\,R_t}} \cdot
\left(  \ds{\fr{R_t}{z}} \right)^{d-\fr{1}{\nu}}, \qq R_t<z<z_1\nonumber\\
\rho_c(z)\q &=&\q \Phi_{\rm sc}(z)/N_t.
\end{eqnarray}
At larger distances $z_1<z<R_B$, each loop carries one side--chain
and thus $\rho_c(z)=z^{-d}$, $\Phi_{\rm sc}(z) = N_t \, z^{-d}$.
The crossover between these two regimes occurs at a distance $z_1$
given by:
\begin{equation}\label{zeta1}
z_1 \q = \q a\;
\left[\ds{\fr{N_B}{N_t^{\nu(d-1)-1}}}\right]^{\nu}.
\end{equation}
For $d=4$ and $\nu=1/2$, Eq.(\ref{zeta1}) gives back the mean field result
of Section \ref{secondo13}, $z_1^2 = N_B/N_t^{1/2}$.

Far from the surface, for $z>R_B\sim a\,N_B^{\nu}$, the
comb--copolymer behaves as an effective linear chain of blobs, the
individual blobs being triblock copolymers with one side--chain
per blob and thus $\rho_c(z)=z^{-d+1/\nu}/N_B$, $\Phi_{\rm sc}(z)
= N_t \, z^{-d+1/\nu}/N_B$.

At short distances, $0<z<R_t$, the density of side chain monomers
is dominated by those side--chains for which the junction point
belongs to the same blob of size $z$. The density of side--chain
monomers is thus equal to the product of the junction point
density $\rho_c$ by the number of monomers of the side chain in
this same blob
 $z^{1/\nu}$, $\Phi_{\rm sc}(z) \sim z^{1/\nu}\rho_c(z)$.
To place a branching point at position $z$ we need to place the
relevant core monomer and to let a tail of size $N_t$ start there.
In contrast to the mean field description, the side--chain is
correlated with the backbone. The branching point density
$\rho_c(z)$ is derived in the Appendix D. It is constructed from
the backbone monomer density, the partition function of a free
tail \cite{SeJo} and  the three leg vertex at the branching point.
We obtain:
\begin{equation}
 \rho_c(z) = {\Gamma\over
N_B R_t}\left({z\over  R_t}\right)^{\alpha}
\end{equation}
with an exponent $\alpha = -d/2-1 +\gamma/(2\nu) +1/\nu -\theta_1\simeq -0.27$.
The junction points are thus weakly localized at the surface where excluded volume
correlations are screened.
The des Cloizeaux exponent $\theta_1=(\sigma_1-\sigma_3)/\nu$ is introduced by the
three leg vertex (see Appendix D).
The side--chain monomer concentration can be deduced as:
\begin{equation}
\Phi_{\rm sc}(z) \q = \q z^{1/\nu}\rho_c(z)= {\Gamma N_t\over N_B
R_t}\left({z\over  R_t}\right)^\beta,
\end{equation}
where the exponent $\beta$ is close to $1.4$.

Summarizing, for $R_t<z_1$, {\it i.e.} in the mushroom regime for
the comb--polymer, the side--chains monomer volume fraction is
given by:
\begin{equation}
\Phi_{\rm sc}(z) \q=\q \left\{
\begin{array}{ll}
\ds{\fr{\Gamma\,N_t}{N_B\,R_t}}\cdot \left(\ds{\fr{z}{R_t}}
\right)^{-d/2-1 +\gamma/(2\nu) +2/\nu -\theta_1} & 0 \leq z \leq R_t,\\
\ds{\fr{\Gamma\,N_t}{N_B\,R_t}}\cdot \left(\ds{\fr{R_t}{z}}
\right)^{d-\fr{1}{\nu}} & R_t \leq z \leq z_1 ,\\
\ds{\fr{N_t}{z^d}} & z_1 \leq z \leq R_B, \\
\ds{\fr{N_t}{N_B}} \; z^{\fr{1}{\nu}-d}&  R_B \leq z \leq
\lambda_p.
\end{array}
\right.
\end{equation}

\subsection{Comb--Copolymer Brush Regime: Scaling}
\label{secondo22}

So far, we have only considered the mushroom limit where the
side--chains do not interact. The number of side chains per unit
area in the proximal layer of thickness $ R_t $ is $\sigma=\Gamma
/N_B$ and the side chains interact if $\sigma R_t^{d-1}>1$; in
this case the side chains stretch and form a polymer brush. This
occurs if $R_t \geq z_1$  or as we have shown in Section
\ref{secondo14}, the adsorbed combs enter the brush regime for:
\begin{equation}
 N_t \geq N_B^{\fr{1}{\nu \,(d-1)}}.
\end{equation}
In the mean field approximation, ($d=4$ and $\nu=1/2$) this
condition gives $ R_t \sim N_B^{1/3} \sim z^*$, in agreement with
our mean field results of Section \ref{secondo12} for the
cross--over from the mushroom to brush configuration. We will
limit our analysis to the study of strong backbone adsorption and
thus we assume that the occurrence of large backbone loops with
many side--chains anchored is negligible.

In the brush regime, the side--chains extend into the bulk from
their anchoring point on the backbone in a sequence of blobs of
size $D_b \sim a\, g^{\nu}$, where $g\sim \sigma^{-1/\nu(d-1)}$ is
the number of monomers per blob. The grafting density is
$\sigma=\Gamma /N_B$ and the blob size is  $D_b \sim a
\,N_B^{1/(d-1)}$. The concentration of side--chain monomers
belonging to the first blob close to the surface is the same as
the concentration in an adsorbed layer of comb--copolymers where
the side chains would have $g$ monomers or a radius $D_b$. It is
obtained from the results of the previous section by replacing
$N_t$ by $g$ and $R_t$ by $D_b$. The cross--over length $z_1$ is
then given by $z_1=a\,N_B^{1/(d-1)} =D_b$ and the density of
side--chain monomers in the brush
\begin{equation}
\Phi_{\rm sc}(z) \q=\q \ds{\fr{\Gamma}{N_B\,z_1^{1-\fr{1}{\nu}}}}
\cdot \left( \ds{\fr{z}{z_1}}
\right)^{\fr{\gamma}{2\,\nu}+\frac{(d-2)}{2}}.
\end{equation}

The thickness of the brush in the blob model is
\begin{equation}\label{L2}
L \;\sim\; a\, N_t \;\sigma^{\fr{1-\nu}{\nu\,(d-1)}}
\sim a \, N_t \, N_B^{-\fr{1-\nu}{\nu\,(d-1)}}.
\end{equation}
The free energy of the side chains is $k_b\,T$ per blob or per
side chain
 \begin{equation}\label{fttt}
\;F_t\,\sim\,\left(\fr{a^2}{k_b\,T} \right)N_t \cdot
\sigma^{1/\nu(d-1)}\sim\,\left(\fr{a^2}{k_b\,T} \right)N_t \cdot
N_B^{-1/\nu(d-1)}
\end{equation}
The adsorption energy of the backbone chains must compensate the
stretching energy of the side--chain brush. This requires an
adsorption energy per monomer $\epsilon \sim F_t/N_B \sim N_t
\cdot N_B^{-(1+1/\nu(d-1))} $. The thickness of the adsorbed
backbone layer is then $\lambda_c \sim a/\epsilon^{\nu}$ or
\begin{equation}
\lambda_c \q\sim\q a\;\ds{  \fr{  N_B^{ \nu+\fr{1}{d-1}}
}{N_t^{\nu}} },
\end{equation}
As the length of the side chains increases and becomes larger than
$N_t\ll N_B^{1/(d-1)\,\nu}$, the thickness of the backbone
adsorbed layer decreases from the radius $R_B$ between two
branching points and the adsorbed polymer amount decreases. The
adsorbed layer is only stable if its thickness is larger than the
proximal distance $b$ introduced in Eq.(\ref{Edw1}). For longer
side chains, there is no adsorption of the comb--copolymer
\begin{equation}
N_t \ge N_B^{1+\frac{1}{\nu(d-1)}},
\end{equation}
where the exponent $1+1/\nu(d-1)$ is equal to $11/6$ for swollen
chains ($d=3$ and $\nu=3/5$) in a good solvent.

\section{Conclusions}\label{terzobubuah}

We propose here a theoretical investigation of the adsorption of
partly adsorbing comb--copolymers, with an adsorbing skeleton and
non--adsorbing side--grafts. Three regimes were identified: a
"mushroom" regime characterized by having an adsorbed backbone
layer on the surface and the side--chains dangling from the
backbone in a swollen state, a "brush" regime in which they
develop a uniform multi--chains coating with an extended layer of
non--adsorbing segments, and a non--adsorbed regime. Depending on
the size of side--chains, size of skeleton length between
side--chains, and adsorption parameters, the scaling laws for the
transitions between the different regimes, the adsorption free
energy and the thickness of adsorbed layers are derived using a
combination of mean field and scaling approaches. In the case of
swollen (ideal) chains, $\nu=3/5$ and $d=3$ ($\nu=1/2$ and $d=4$),
the threshold between mushroom and brush configurations and the
desorption threshold can be expressed as $N_t > N_B^{5/6}$ ($N_t >
N_B^{2/3}$) and $N_t \sim N_B^{11/6}$  ($N_t \sim N_B^{5/3}$). In
the brush regime we find that the thickness of the backbone
adsorbed layer and the vertical extension of the brush for a
swollen (ideal) chain scale as $\lambda_c \sim a \, N_B^{11/10} \,
N_t^{-3/5}$ ($\lambda_c \sim a \, N_B^{5/6} \, N_t^{-1/2}$) and
$L\sim a \, N_t \, N_B^{-1/3}$ ($L\sim a \, N_t \, N_B^{-1/3}$),
respectively. These predictions could be checked quantitatively,
by experiments able to investigate polymer adsorption, such as
Neutron or X--ray reflectometry, ellipsometry, quartz
microbalance, or Surface Force Apparatus, and work is currently in
progress in our group in this direction. Qualitatively, this new
family of copolymers can lead to rather thick layers, without the
difficulty often encountered when trying to prepare long
conventional ({\it i.e.} diblock or triblock) copolymers. These
multiblock copolymers may thus be interesting in numerous
applications, in which the adsorption of rather large objects
(proteins, cells) should be prevented, or controlled. They already
demonstrated very interesting performances in the context of DNA
sequencing and capillary electrophoresis. In this application, the
interesting regime is probably the "brush" regime, because a
uniform layer with no access of the analytes to the wall is
wanted. The extension of the brush leads to thicker layers, which
should be favorable, but also smaller adsorption free energies, so
that a compromise has to be found. Probably, a "weakly extended"
brush is a good aim on the practical side. An important aspect of
the problem, on the practical side, is the adsorption kinetics. It
has been well recognized that the adsorption of large polymers on
surfaces is strongly constrained by the kinetics of penetration of
a new polymers across the already adsorbed layer, so that the
thermodynamic equilibrium, which is discussed in the present
article, can be hard to reach. The adsorption kinetics of high
molecular weight polymers is a very difficult problem on the
theoretical side, and it is beyond the scope of the present
article. We believe, however, that numerous information useful for
experimental development of applications can be gained from the
present approach. In particular, there are practical ways to
minimize kinetic barriers to adsorption, {\it e.g.} by performing
adsorption from a semi--dilute, rather than dilute, solution.\\ \\
Acknowledgments: A.S. acknowledges a "Marie Curie" Postdoctoral
fellowship from the EU (HPMF-CT-2000-00940) and thanks Dr. P. Sens
for very valuable discussions. This work was partly supported by a
grant from Association pour la Recherche sur le Cancer (ARC).

\section*{Appendix}\label{terzobubu}

\subsection{Triblock copolymer: Monomer Volume Fraction}\label{terzobusZ}
\begin{equation}
\begin{array}{lll}
\Phi_{\rm \ell}(z)\q=\q\left\{
\begin{array}{lll}
 2& (z+d)^{-2}  &\q 0 \leq z \leq z^*, \\
z^{*^6}  & z^{-8}  &\q z^* \leq z \leq \lambda,
\end{array}
\right.
\\ \\
\Phi_{\rm t}(z)\q=\q\left\{
\begin{array}{ll}
 \ds{\fr{z}{N_B}} \,
\ln\left[\fr{z^*}{z} \right] & 0 < z < z^*, \\
  z^{-2} & z^*< z < \lambda,
\end{array}
\right.
\\ \\
\Phi^{\rm TB}_{\rm sc}(z) \q=\q \left\{
\begin{array}{lll}
\ds{\fr{\Gamma_o}{N_B\,R_t}} & \q z^2   & \q d<z<R_t,\\
\ds{\fr{\Gamma_o\,R_t^3}{N_B}} & \q z^{-2}  & \q R_t<z<z^*,\\
\Gamma_o\,R_t^3\,N_B  & \q z^{-8} & \q z^*<z<\lambda,
\end{array}
\right.
\end{array}
\end{equation}
where $z^*=(N\, b\,\log({N/b^2}))^{1/3}$.
\begin{equation}
\Phi^{\rm TB}(z)\;=\;\left\{
\begin{array}{ll}
2\; (z+d)^{-2} \,+\, \fr{\Gamma_o}{N_B\,R_t}\cdot z^4 \,+\,
\ds{\fr{z}{N_B}}  \; \ln\left(  \ds{\fr{z^*}{z}} \right)
& \q 0 \leq z \leq R_t, \\ \\
\left[ 2\,+\, \ds{\fr{\Gamma_o\,R_t^3}{N_B}}\right]\; z^{-2} \;+\;
\ds{\fr{z}{N_B}}  \; \ln\left(  \ds{\fr{z^*}{z}} \right)
& \q R_t \leq z \leq z^*, \\ \\
 z^{-2}  \;+\; \left[ z^{*^6}\;+\; \Gamma_o\,R_t^3\,N_B  \right]\cdot
z^{-8} & \q z^* \leq z \leq \lambda
\end{array}
\right.
\end{equation}

\subsection{Comb--Copolymer: Side--Chains Monomer Volume Fraction}\label{terzobusZ2}

\begin{equation}
\Phi^{\rm co}_{\rm sc}(z) \q=\q \left\{
\begin{array}{lll}
\ds{\fr{\Gamma_o}{N_B\,R_t}} \;z^2 & 0<z<R_t\\
\ds{\fr{\Gamma_o\,R_t^3}{N_B}\;z^{-2}} & R_t<z<z_1\\
\Gamma_o\,N_t\;z^{-4}& z_1<z<R_B\\
\Gamma_o\,\ds{\fr{N_t}{N_B}}\; z^{-2}& R_B<z<\lambda_p
\end{array}
\right.
\end{equation}

\subsection{Propagator and Partition Function of a Side--chain in
a Triblock Copolymer Adsorbed Layer}
\label{johner1}
\subsubsection{Chain Propagator}

We calculate here the propagator of a side chain in the triblock
copolymer adsorbed layer between the junction point at coordinate
$z_c$ and and the free end point at coordinate $z$,
$Z_t^f(z,N_t\vert z_c)$. The Laplace transform of this propagator
$\tilde Z_t^f(z,p\vert z_c)=\int_0^\infty dN_t \exp -(p\,N_t)\;
Z_t^f(z,N_t\vert z_c)$
\begin{equation}
-\delta(z-z_c) \q=\q {\partial^2 \tilde Z_t^f(z,p\vert
z_c)\over\partial z^2} \,-\,(p +{2\over z^2})\;\tilde
Z_t^f(z,p\vert z_c),
\label{edwards}
 \end{equation}
with the boundary condition that it vanishes at the wall $z=0$.
The solution of this equation is
\begin{equation}
\tilde Z_t^f(z,p\vert z_c) \q=\q f_-(z_<)\; f_+(z_>),
\label{propagator}
\end{equation}
where we have defined $z_{<,>}=\mbox{ min,max}(z_c,z)$. The functions $f_+$ and $f_-$ are
given  by
\begin{eqnarray}
f_+(z,p) &=& \sqrt{\pi\over 2p}\;e^{-\sqrt p z}\;\left(1+{1\over z\sqrt p}\right) \nonumber, \\
f_-(z,p) &=& \sqrt{2\over \pi} \;\left(\cosh(\sqrt p
z)-{\sinh(\sqrt p z)\over \sqrt p z}\right).
\end{eqnarray}
The following asymptotic limits are useful:
\begin{eqnarray}
z_>&\ll& R_t\quad \tilde Z_t^f(z,p=0\vert z_c) = {z_<^2\over 3 z_>},\\
z_<&\gg& R_t\quad \tilde Z_t^f(z,p\vert z_c) = {1\over 2\sqrt
p}e^{-\sqrt p (z_>-z_<)} \; \lim_{p\rightarrow\infty}{1\over p}\;
\delta(z-z_c). \label{propagatorasympt.}
\end{eqnarray}

\subsubsection{Partition Function}

The partition function of a side--chain $Z_t^f(z_c)$ with the
junction point at position  $z_c$ is obtained by integration of
$\tilde Z_t^f(z,p\vert z_c)$:
\begin{equation}
\tilde Z_t^f(z_c,p) = f_-(z_c,p)\int_{z_c}^\infty f_+(z,p)\;{\rm
d}z  +  f_+(z_c,p)\int_{0}^{z_c} f_-(z,p)\;{\rm d}z.
\label{fixedend}
\end{equation}

At short distances from the wall $z_c\ll R_t$, the partition function is dominated
by the contribution to the integral coming from  $z>z_c$:
\begin{eqnarray}
\tilde Z_t ^f(z_c,p) &=& -{z_c^2\over 3}\log(\sqrt p z_c),\\
Z_t ^f(z_c,N_t) &=&{ z_c^2\over 6 N_t}. \label{ZT}
\end{eqnarray}
At larger distances from the wall, $z_c\gg R_t$
the two integrals are equal to  $1/2p$ and $Z(z_c,N_t)=1$;
the side chains are almost free chains.

The density of copolymer junction points $\rho_{c}^{\rm TB}(z)$
and the side--chain monomers volume fraction $\Phi_{sc}^{\rm TB}$
can be calculated using these more precise values of the
propagator and partition function; one finds the following
results:

\begin{equation}
\begin{array}{lll}
\rho_c^{\rm TB}(z)\q=\q \left\{
\begin{array}{ll}
\ds{{1\over  N_B\, R_t}}         & \q z\ll R_t \nonumber ,\\ \\
\ds{{1\over  N_B}{R_t\over z^2} }& \q R_t\ll z\ll z^\star,
\end{array}
\right.
\\ \\
\Phi_{sc}^{\rm TB}(z) \q=\q \left\{
\begin{array}{ll}
\rho^{\rm TB}_c(z<R_t)\cdot z^2 &\quad z\ll R_t \nonumber, \\
 \rho^{\rm TB}_c(z) \cdot N_t &\quad z\gg R_t.
\end{array}
\right.
\end{array}
\end{equation}

These results are similar to those obtained in the main text.

\subsection{Density of Comb--Copolymer Branching Points}
\label{johner2}

In order to determine using scaling arguments the density of
branching points of the comb--copolymer at a distance $z_c$
smaller than the side chain radius, we first calculate the
partition function $Z^f_t(z_c, R_t)$ of a side chain with the
branching point at position $z_c$ and radius $R_t$.

In the limit where $z_c$ is of the order of a monomer size $a$,
the side--chain behaves as a tail in the adsorbed layer and
$Z^f_t(a, R_t) \sim R_t^{\frac{\gamma-\nu(d-2)}{2\nu}-\frac
1{\nu}}$.

We now consider the case where $z_c = R_t$. The probability to
find a side chain at a distance $z_c$ is proportional to the local
monomer concentration $c(z_c)\sim z_c^{1/\nu-d}$. If the side
chain is not connected to the backbone but is free, its partition
function is $Z^o\sim z_c^{(\gamma-1)/\nu}$. The partition function
of the side chain also contains a factor associated to the
branching point. This is best written in terms of the vertex
exponents introduced by Duplantier \cite{duplantier}. In the
partition function of a branched polymer chain, each vertex having
$k$ legs is associated, for a chain of size $z_c$, to a factor
$z_c^{\sigma_k/\nu}$ where $\sigma_k$ is the corresponding vertex
exponent. The formation of a branching  point in the
comb--copolymer corresponds to the disappearance of a two legs
vertex on the backbone and a one leg vertex (the side--chain free
end) and the appearance of a three legs vertex (the branching
point); it is therefore associated to a weight
$z_c^{(\sigma_3-\sigma_1 -\sigma_2)/\nu}$. Note that it is
sufficient to consider that the backbone chain has a size $z_c$
since, in an adsorbed polymer layer, the local screening length is
the distance to the adsorbing surface $z_c$. Considering all these
factors, we obtain the partition function
\begin{equation}
Z^f_t(z_c, R_t) \sim
z_c^{(1-d\nu+\gamma-1+\sigma_3-\sigma_2-\sigma_1)/\nu} \sim
z_c^{\gamma/\nu-d-\theta_1}.
\end{equation}
We have here used the fact that $\sigma_2=0$ and introduced the
contact exponent $\theta_1=(\sigma_1-\sigma_3)/\nu \simeq 0.45$
first considered by Des Cloizeaux \cite{descloizeaux}.

The partition function $Z^f_t(z_c, R_t)$ is obtained by a
scaling law extrapolating between these two asymptotic limits
\begin{equation}
Z^f_t(z_c, R_t) \sim z_c^{\alpha}
R_t^{\frac{\gamma-\nu(d-2)}{2\nu}-\frac 1{\nu}},
\end{equation}
with an exponent $\alpha = -d/2-1 +\gamma/2\nu +1/\nu -\theta_1\simeq -0.27$.

The density of branching points is proportional to this partition
function; the pre--factor is obtained by imposing that the total
number  of branching points per unit area over a thickness $R_t$
is of order $\Gamma/N_B$:
\begin{equation}
\rho_c(z_c) = {\Gamma\over N_B R_t}{\left(z_c\over
R_t\right)}^{\alpha}
\end{equation}

\newpage

\bibliography{refs}

\begin{thebibliography}{29}
\expandafter\ifx\csname natexlab\endcsname\relax\def\natexlab#1{#1}\fi
\expandafter\ifx\csname bibnamefont\endcsname\relax
  \def\bibnamefont#1{#1}\fi
\expandafter\ifx\csname bibfnamefont\endcsname\relax
  \def\bibfnamefont#1{#1}\fi
\expandafter\ifx\csname citenamefont\endcsname\relax
  \def\citenamefont#1{#1}\fi
\expandafter\ifx\csname url\endcsname\relax
  \def\url#1{\texttt{#1}}\fi
\expandafter\ifx\csname urlprefix\endcsname\relax\def\urlprefix{URL }\fi
\providecommand{\bibinfo}[2]{#2}
\providecommand{\eprint}[2][]{\url{#2}}

\bibitem[{\citenamefont{Hjerten and Kubo}(2000)}]{Hjerten}
\bibinfo{author}{\bibfnamefont{S.}~\bibnamefont{Hjerten}} \bibnamefont{and}
  \bibinfo{author}{\bibfnamefont{K.}~\bibnamefont{Kubo}},
  \bibinfo{journal}{Electrophoresis} \textbf{\bibinfo{volume}{14}},
  \bibinfo{pages}{390} (\bibinfo{year}{2000}).

\bibitem[{\citenamefont{Marshall and Pennisi}(1998)}]{1}
\bibinfo{author}{\bibfnamefont{E.}~\bibnamefont{Marshall}} \bibnamefont{and}
  \bibinfo{author}{\bibfnamefont{E.}~\bibnamefont{Pennisi}},
  \bibinfo{journal}{Science} \textbf{\bibinfo{volume}{280}},
  \bibinfo{pages}{994} (\bibinfo{year}{1998}).

\bibitem[{\citenamefont{Salas-Solano et~al.}(2000)\citenamefont{Salas-Solano,
  Schmalzing, Koutny, Buonocore, Adourian, Matsudaira, and Ehrlich}}]{5}
\bibinfo{author}{\bibfnamefont{O.}~\bibnamefont{Salas-Solano}},
  \bibinfo{author}{\bibfnamefont{D.}~\bibnamefont{Schmalzing}},
  \bibinfo{author}{\bibfnamefont{L.}~\bibnamefont{Koutny}},
  \bibinfo{author}{\bibfnamefont{S.}~\bibnamefont{Buonocore}},
  \bibinfo{author}{\bibfnamefont{A.}~\bibnamefont{Adourian}},
  \bibinfo{author}{\bibfnamefont{P.}~\bibnamefont{Matsudaira}},
  \bibnamefont{and} \bibinfo{author}{\bibfnamefont{D.}~\bibnamefont{Ehrlich}},
  \bibinfo{journal}{Anal. Chem.} \textbf{\bibinfo{volume}{72}},
  \bibinfo{pages}{3129} (\bibinfo{year}{2000}).

\bibitem[{\citenamefont{Carrilho}(2000)}]{14}
\bibinfo{author}{\bibfnamefont{E.}~\bibnamefont{Carrilho}},
  \bibinfo{journal}{Electrophoresis} \textbf{\bibinfo{volume}{21}},
  \bibinfo{pages}{55} (\bibinfo{year}{2000}).

\bibitem[{\citenamefont{Viovy}(2000)}]{JLV}
\bibinfo{author}{\bibfnamefont{J.-L.} \bibnamefont{Viovy}},
  \bibinfo{journal}{Rev. Mod. Phys.} \textbf{\bibinfo{volume}{72}},
  \bibinfo{pages}{813} (\bibinfo{year}{2000}).

\bibitem[{\citenamefont{Ajdari}(1995)}]{Ajdari}
\bibinfo{author}{\bibfnamefont{A.}~\bibnamefont{Ajdari}},
  \bibinfo{journal}{Phys. Rev. Lett.} \textbf{\bibinfo{volume}{75}},
  \bibinfo{pages}{755} (\bibinfo{year}{1995}).

\bibitem[{\citenamefont{Horvath and Dolnik}(2001)}]{16}
\bibinfo{author}{\bibfnamefont{J.}~\bibnamefont{Horvath}} \bibnamefont{and}
  \bibinfo{author}{\bibfnamefont{V.}~\bibnamefont{Dolnik}},
  \bibinfo{journal}{Electrophoresis} \textbf{\bibinfo{volume}{22}},
  \bibinfo{pages}{644} (\bibinfo{year}{2001}).

\bibitem[{\citenamefont{Righetti et~al.}(2001)\citenamefont{Righetti, Gelfi,
  Verzola, and Castelletti}}]{17}
\bibinfo{author}{\bibfnamefont{P.~G.} \bibnamefont{Righetti}},
  \bibinfo{author}{\bibfnamefont{C.}~\bibnamefont{Gelfi}},
  \bibinfo{author}{\bibfnamefont{B.}~\bibnamefont{Verzola}}, \bibnamefont{and}
  \bibinfo{author}{\bibfnamefont{L.}~\bibnamefont{Castelletti}},
  \bibinfo{journal}{Electrophoresis} \textbf{\bibinfo{volume}{22}},
  \bibinfo{pages}{603} (\bibinfo{year}{2001}).

\bibitem[{\citenamefont{Doherty et~al.}(2002)\citenamefont{Doherty, Berglund,
  Buchholz, Kourkine, Przybycien, Tilton, and Barron}}]{18}
\bibinfo{author}{\bibfnamefont{E.~A.} \bibnamefont{Doherty}},
  \bibinfo{author}{\bibfnamefont{K.~D.} \bibnamefont{Berglund}},
  \bibinfo{author}{\bibfnamefont{B.~A.} \bibnamefont{Buchholz}},
  \bibinfo{author}{\bibfnamefont{I.~V.} \bibnamefont{Kourkine}},
  \bibinfo{author}{\bibfnamefont{T.~M.} \bibnamefont{Przybycien}},
  \bibinfo{author}{\bibfnamefont{R.~D.} \bibnamefont{Tilton}},
  \bibnamefont{and} \bibinfo{author}{\bibfnamefont{A.~E.}
  \bibnamefont{Barron}}, \bibinfo{journal}{Electrophoresis}
  \textbf{\bibinfo{volume}{23}}, \bibinfo{pages}{2766} (\bibinfo{year}{2002}).

\bibitem[{\citenamefont{Liang et~al.}(2001)\citenamefont{Liang, Liu, Song, and
  Chu}}]{37}
\bibinfo{author}{\bibfnamefont{D.}~\bibnamefont{Liang}},
  \bibinfo{author}{\bibfnamefont{T.}~\bibnamefont{Liu}},
  \bibinfo{author}{\bibfnamefont{L.}~\bibnamefont{Song}}, \bibnamefont{and}
  \bibinfo{author}{\bibfnamefont{B.}~\bibnamefont{Chu}}, \bibinfo{journal}{J.
  Chrom.} \textbf{\bibinfo{volume}{909}}, \bibinfo{pages}{271}
  (\bibinfo{year}{2001}).

\bibitem[{\citenamefont{Liang and Chu}(1998)}]{38}
\bibinfo{author}{\bibfnamefont{D.}~\bibnamefont{Liang}} \bibnamefont{and}
  \bibinfo{author}{\bibfnamefont{B.}~\bibnamefont{Chu}},
  \bibinfo{journal}{Electrophoresis} \textbf{\bibinfo{volume}{19}},
  \bibinfo{pages}{2447} (\bibinfo{year}{1998}).

\bibitem[{\citenamefont{Madabhushi}(1998)}]{12}
\bibinfo{author}{\bibfnamefont{R.~S.} \bibnamefont{Madabhushi}},
  \bibinfo{journal}{Electrophoresis} \textbf{\bibinfo{volume}{19}},
  \bibinfo{pages}{224} (\bibinfo{year}{1998}).

\bibitem[{\citenamefont{Chiari et~al.}(2000)\citenamefont{Chiari, Cretich, and
  Horvath}}]{24}
\bibinfo{author}{\bibfnamefont{M.}~\bibnamefont{Chiari}},
  \bibinfo{author}{\bibfnamefont{M.}~\bibnamefont{Cretich}}, \bibnamefont{and}
  \bibinfo{author}{\bibfnamefont{J.}~\bibnamefont{Horvath}},
  \bibinfo{journal}{Electrophoresis} \textbf{\bibinfo{volume}{21}},
  \bibinfo{pages}{1521 } (\bibinfo{year}{2000}).

\bibitem[{\citenamefont{Albarghouthi et~al.}(2001)\citenamefont{Albarghouthi,
  Buchholz, Doherty, Bogdan, Zhou, and Barron}}]{22}
\bibinfo{author}{\bibfnamefont{M.~N.} \bibnamefont{Albarghouthi}},
  \bibinfo{author}{\bibfnamefont{B.~A.} \bibnamefont{Buchholz}},
  \bibinfo{author}{\bibfnamefont{E.~A.} \bibnamefont{Doherty}},
  \bibinfo{author}{\bibfnamefont{F.~M.} \bibnamefont{Bogdan}},
  \bibinfo{author}{\bibfnamefont{H.}~\bibnamefont{Zhou}}, \bibnamefont{and}
  \bibinfo{author}{\bibfnamefont{A.~E.} \bibnamefont{Barron}},
  \bibinfo{journal}{Electrophoresis} \textbf{\bibinfo{volume}{22}},
  \bibinfo{pages}{737} (\bibinfo{year}{2001}).

\bibitem[{\citenamefont{Zhou et~al.}(2000)\citenamefont{Zhou, Miller, Sosic,
  Buchholz, Barron, Kotler, and Karger}}]{6}
\bibinfo{author}{\bibfnamefont{H.}~\bibnamefont{Zhou}},
  \bibinfo{author}{\bibfnamefont{A.~W.} \bibnamefont{Miller}},
  \bibinfo{author}{\bibfnamefont{Z.}~\bibnamefont{Sosic}},
  \bibinfo{author}{\bibfnamefont{B.}~\bibnamefont{Buchholz}},
  \bibinfo{author}{\bibfnamefont{A.~E.} \bibnamefont{Barron}},
  \bibinfo{author}{\bibfnamefont{L.}~\bibnamefont{Kotler}}, \bibnamefont{and}
  \bibinfo{author}{\bibfnamefont{B.~L.} \bibnamefont{Karger}},
  \bibinfo{journal}{Anal. Chem.} \textbf{\bibinfo{volume}{72}},
  \bibinfo{pages}{1045} (\bibinfo{year}{2000}).

\bibitem[{\citenamefont{Barbier et~al.}(2002)\citenamefont{Barbier, Buchholz,
  Barron, and Viovy}}]{26}
\bibinfo{author}{\bibfnamefont{V.}~\bibnamefont{Barbier}},
  \bibinfo{author}{\bibfnamefont{B.~A.} \bibnamefont{Buchholz}},
  \bibinfo{author}{\bibfnamefont{A.~E.} \bibnamefont{Barron}},
  \bibnamefont{and} \bibinfo{author}{\bibfnamefont{J.-L.} \bibnamefont{Viovy}},
  \bibinfo{journal}{Electrophoresis} \textbf{\bibinfo{volume}{23}},
  \bibinfo{pages}{1441} (\bibinfo{year}{2002}).

\bibitem[{\citenamefont{Joanny and Johner}(1996)}]{JoaJoh1}
\bibinfo{author}{\bibfnamefont{J.-F.} \bibnamefont{Joanny}} \bibnamefont{and}
  \bibinfo{author}{\bibfnamefont{A.}~\bibnamefont{Johner}},
  \bibinfo{journal}{J. Phys. II France} \textbf{\bibinfo{volume}{6}},
  \bibinfo{pages}{511} (\bibinfo{year}{1996}).

\bibitem[{\citenamefont{Semenov et~al.}(1996)\citenamefont{Semenov,
  Bonet-Avalos, Johner, and Joanny}}]{SeBoJohJoa}
\bibinfo{author}{\bibfnamefont{A.~N.} \bibnamefont{Semenov}},
  \bibinfo{author}{\bibfnamefont{J.}~\bibnamefont{Bonet-Avalos}},
  \bibinfo{author}{\bibfnamefont{A.}~\bibnamefont{Johner}}, \bibnamefont{and}
  \bibinfo{author}{\bibfnamefont{J.-F.} \bibnamefont{Joanny}},
  \bibinfo{journal}{Macromolecules} \textbf{\bibinfo{volume}{29}},
  \bibinfo{pages}{2179} (\bibinfo{year}{1996}).

\bibitem[{\citenamefont{Semenov et~al.}(1998)\citenamefont{Semenov, Joanny, and
  Johner}}]{SeJoaJoh}
\bibinfo{author}{\bibfnamefont{A.~N.} \bibnamefont{Semenov}},
  \bibinfo{author}{\bibfnamefont{J.-F.} \bibnamefont{Joanny}},
  \bibnamefont{and} \bibinfo{author}{\bibfnamefont{A.}~\bibnamefont{Johner}},
  in \emph{\bibinfo{booktitle}{Theoretical and Mathematical Models in Polymer
  Research}} (\bibinfo{publisher}{Academic Press}, \bibinfo{year}{1998}), pp.
  \bibinfo{pages}{37--82}.

\bibitem[{\citenamefont{Marques and Joanny}(1989)}]{MaJoa1}
\bibinfo{author}{\bibfnamefont{C.~M.} \bibnamefont{Marques}} \bibnamefont{and}
  \bibinfo{author}{\bibfnamefont{J.-F.} \bibnamefont{Joanny}},
  \bibinfo{journal}{Macromolecules} \textbf{\bibinfo{volume}{22}},
  \bibinfo{pages}{1454} (\bibinfo{year}{1989}).

\bibitem[{\citenamefont{Johner and Joanny}(1997)}]{JohJoa}
\bibinfo{author}{\bibfnamefont{A.}~\bibnamefont{Johner}} \bibnamefont{and}
  \bibinfo{author}{\bibfnamefont{J.-F.} \bibnamefont{Joanny}},
  \bibinfo{journal}{Macromol. Theory Simul.} \textbf{\bibinfo{volume}{6}},
  \bibinfo{pages}{479} (\bibinfo{year}{1997}).

\bibitem[{\citenamefont{Gennes}(1981)}]{DeGads1}
\bibinfo{author}{\bibfnamefont{P.~G.~D.} \bibnamefont{Gennes}},
  \bibinfo{journal}{Macromolecules} \textbf{\bibinfo{volume}{14}},
  \bibinfo{pages}{1637} (\bibinfo{year}{1981}).

\bibitem[{\citenamefont{de~Gennes}(1980)}]{DeGads2}
\bibinfo{author}{\bibfnamefont{P.~G.} \bibnamefont{de~Gennes}},
  \bibinfo{journal}{Macromolecules} \textbf{\bibinfo{volume}{13}},
  \bibinfo{pages}{1069} (\bibinfo{year}{1980}).

\bibitem[{\citenamefont{Marques and Joanny}(1990)}]{MaJoa}
\bibinfo{author}{\bibfnamefont{C.~M.} \bibnamefont{Marques}} \bibnamefont{and}
  \bibinfo{author}{\bibfnamefont{J.-F.} \bibnamefont{Joanny}},
  \bibinfo{journal}{Macromolecules} \textbf{\bibinfo{volume}{23}},
  \bibinfo{pages}{268} (\bibinfo{year}{1990}).

\bibitem[{\citenamefont{Marques et~al.}(1988)\citenamefont{Marques, Joanny, and
  Leibler}}]{MaJoaLei}
\bibinfo{author}{\bibfnamefont{C.~M.} \bibnamefont{Marques}},
  \bibinfo{author}{\bibfnamefont{J.-F.} \bibnamefont{Joanny}},
  \bibnamefont{and} \bibinfo{author}{\bibfnamefont{L.}~\bibnamefont{Leibler}},
  \bibinfo{journal}{Macromolecules} \textbf{\bibinfo{volume}{21}},
  \bibinfo{pages}{1051} (\bibinfo{year}{1988}).

\bibitem[{\citenamefont{Grosberg and Khoklov}(1994)}]{Grosberg}
\bibinfo{author}{\bibfnamefont{A.~Y.} \bibnamefont{Grosberg}} \bibnamefont{and}
  \bibinfo{author}{\bibfnamefont{A.~R.} \bibnamefont{Khoklov}},
  \emph{\bibinfo{title}{Statistical Physics of Macromolecules}}
  (\bibinfo{publisher}{AIP Press, New York}, \bibinfo{year}{1994}).

\bibitem[{\citenamefont{Semenov and Joanny}(1995)}]{SeJo}
\bibinfo{author}{\bibfnamefont{A.~N.} \bibnamefont{Semenov}} \bibnamefont{and}
  \bibinfo{author}{\bibfnamefont{J.~F.} \bibnamefont{Joanny}},
  \bibinfo{journal}{Europhys. Lett.} \textbf{\bibinfo{volume}{29}},
  \bibinfo{pages}{279} (\bibinfo{year}{1995}).

\bibitem[{\citenamefont{Duplantier}(1989)}]{duplantier}
\bibinfo{author}{\bibfnamefont{B.}~\bibnamefont{Duplantier}},
  \bibinfo{journal}{J.Stat.Phys.} \textbf{\bibinfo{volume}{54}},
  \bibinfo{pages}{581} (\bibinfo{year}{1989}).

\bibitem[{\citenamefont{Cloizeaux and Janink}(1987)}]{descloizeaux}
\bibinfo{author}{\bibfnamefont{J.~D.} \bibnamefont{Cloizeaux}}
  \bibnamefont{and} \bibinfo{author}{\bibfnamefont{G.}~\bibnamefont{Janink}},
  \emph{\bibinfo{title}{Les Polym\`eres en Solution: leur mod\'elisation et
  leur structure}} (\bibinfo{publisher}{Editions de Physique},
  \bibinfo{year}{1987}).

\end{thebibliography}

\clearpage

\begin{table}[h]
\begin{center}
\begin{tabular}{|c||l|}
\hline
$\Phi_b$            & \hspace{0.1cm} Bulk monomer volume fraction\\
$N=p\,(N_B \,+\,N_t)$& \hspace{0.1cm} Total number of monomers per comb--polymer\\
$p$                 & \hspace{0.1cm} Number of triblock sub--units per comb--polymer \\
$N_B$               & \hspace{0.1cm} Number of backbone monomers per triblock\\
$N_t$               & \hspace{0.1cm} Number of side--chain monomers per triblock \\
a                   & \hspace{0.1cm} Monomer size ($\sim$ {\rm nm})\\
$\epsilon_o$        & \hspace{0.1cm} adsorption energy per monomer
of a linear chain \\
$\epsilon$          & \hspace{0.1cm} adsorption energy per monomer
of a comb--polymer \\
 b                   & \hspace{0.1cm} Extrapolation length
($\sim 1/\e_o$ for adsorbing linear chains)\\
$R_t= a \sqrt{N_t}$ & \hspace{0.1cm} Side--chain gyration radius\\
$R_B= a \sqrt{N_B}$ & \hspace{0.1cm} gyration radius of backbone
                                     monomers for a triblock \\
$z^*$               & \hspace{0.1cm} Thickness of loops dominated layer over the adsorbing surface\\
$\lambda$           & \hspace{0.1cm} Total thickness of the adsorbed layer\\
$Z^o_{N_B}(z)$      & \hspace{0.1cm} Partition function of a
                                     linear backbone adsorbing chain made of
                                     $N_B$  \\
                    & \hspace{0.1cm} monomers and with one end at position $z$ above the adsorbing surface \\
$Z_N^f(z)$          & \hspace{0.1cm} Partition function of a
                                     non--adsorbed linear chain of N monomers\\
$Z_N(z)$            & \hspace{0.1cm} Partition function of a triblock made of N monomers and \\
                    & \hspace{0.1cm} with one end at position $z$ above the adsorbing surface \\
$\G_o$              & \hspace{0.1cm} Surface coverage (number of monomers per unit surface) for a\\
                    & \hspace{0.1cm} linear chain \\
$\G$                & \hspace{0.1cm} Surface coverage (number of monomers per unit surface) for a\\
                    & \hspace{0.1cm} comb--polymer \\
$L$                 & \hspace{0.1cm} Side--chains brush vertical extension \\
$\sigma$            & \hspace{0.1cm} Surface grafting density \\
$d$                 & \hspace{0.1cm} Space dimension \\

\hline
\end{tabular}
\caption{\label{tab:symbols} Table of Symbols.}
\end{center}
\end{table}
\clearpage

\newpage

\newpage

\begin{figure}[ht]
\vspace{.2in} \centerline {
\includegraphics[width=0.6\textwidth]
{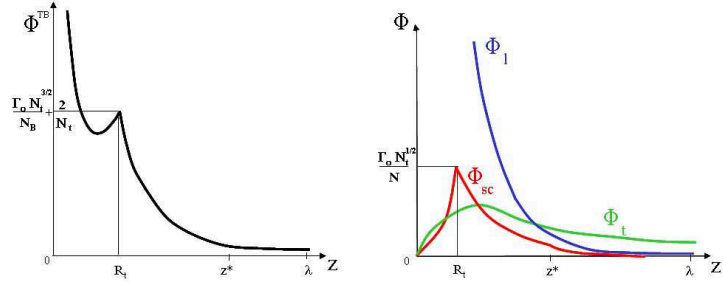}} \caption{\label{fig:4} {\small Triblock's monomer
volume fraction $\Phi^{\rm TB}_{\rm sc} =
\Phi_{\ell}+\Phi_{t}+\Phi_{\rm sc}$: loops ($\Phi_{\ell}(z)$),
tails ($\Phi_{t}(z)$) and side--chains ($\Phi_{\rm sc}(z)$)
contributions.}}
\end{figure}

\newpage

\begin{figure}[ht]\hspace{0cm}
\vspace{.2in}
\centerline {
\includegraphics[width=0.65\textwidth]
{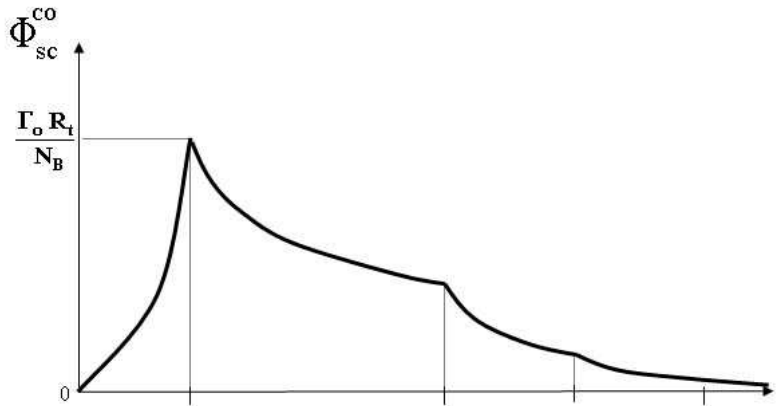}}\caption{\label{fig:6} {\small Combs side--chains
monomer volume fraction, $\Phi^{\rm co}_{\rm sc}(z)$.}}
\end{figure}

\newpage

\begin{figure}[ht]\hspace{0cm}
\vspace{.2in} \centerline {
\includegraphics[width=0.65\textwidth]
{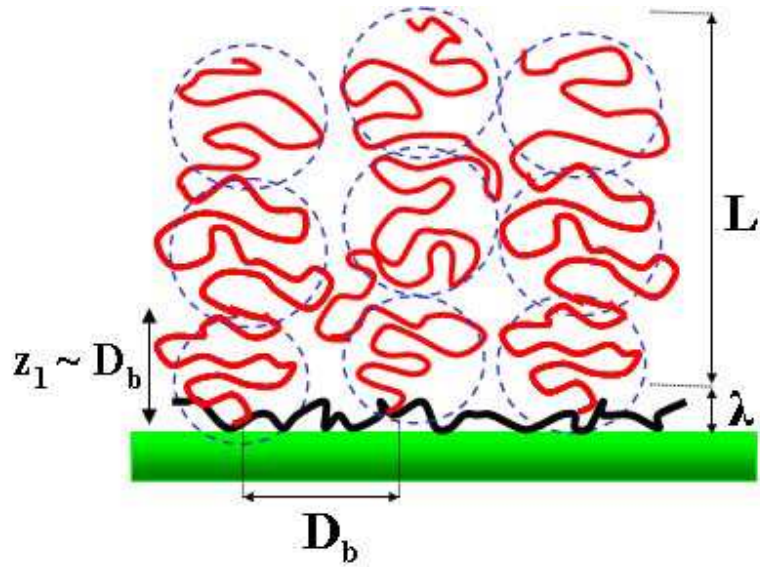}}\caption{\label{fig:7} {\small Brush regime for a
comb--polymer with an adsorbing backbone made of $p$ blocks of
$N_B$ monomers each and with $p$ non--adsorbing side--chains of
$N_t$ monomers each, with $N_t \gg N_B$.}}
\end{figure}

\end{document}